\documentclass{appolb}
\usepackage{placeins}
\usepackage{amsmath}
\usepackage{amssymb}
\usepackage{amsbsy}
\usepackage{amsfonts}
\usepackage{graphicx}
\usepackage{cancel}
\usepackage{soul}

\usepackage[T1]{fontenc}
\usepackage[]{hyperref}
\usepackage[numbers,sort&compress]{natbib}
\usepackage[]{changes} 
\hypersetup{
  unicode=false,          
  pdftoolbar=true,        
 pdfmenubar=true,        
 pdffitwindow=true,     
 pdfstartview={FitH},    
 pdfsubject={GUT and Topological defects},   
 pdfnewwindow=true,      
 pdfcreator={RevTeX},
 colorlinks=true,       
 linkcolor=blue,
 citecolor=blue,        
 filecolor=black,      
 urlcolor=blue,           
  }

\begin{document}
\title{Constraining Neutrino Mixing Schemes \\ with Correlations of Oscillation Data 
\thanks{Presented by S.~Zięba at the XXX Cracow EPIPHANY Conference on Precision Physics at High Energy Colliders, Kraków, January 8-12, 2024}%
}

\author{Biswajit Karmakar, Szymon Zięba
\address{Institute of Physics, University of Silesia,  Katowice, Poland}
}

\maketitle
\begin{abstract}
Correlations obtained from neutrino oscillation data on mixing parameters may help to validate neutrino mixing schemes. In this context, we explore how correlations of neutrino oscillation parameters affect the $\rm{TM}_1$ and $\rm{TM}_2$ mixing scenarios.
\end{abstract}
  
\section{Introduction}

Neutrino oscillation data implies that three types of known flavor neutrinos $|\nu_\alpha \rangle$, ($\alpha =e,\mu,\tau$) mix, in a minimal setup, with three massive states $|\nu_i\rangle$, $(i=1,2,3$), i.e. $|\nu_\alpha \rangle = {\rm U}_{\rm PMNS} |\nu_i\rangle$.
The standard parametrization of the unitary Pontecorvo–Maki–Nakagawa–Sakata (PMNS) mixing matrix is ~\cite{Pontecorvo:1957qd,Maki:1962mu, Kobayashi:1973fv}
\begin{eqnarray*}
{\rm U}_{\rm PMNS} =
\begin{bmatrix}
    c_{12}c_{13} & s_{12}c_{13} & s_{13} e^{i\delta_{\rm CP}} \\
    s_{12}c_{23} -c_{12} s_{13} s_{23} e^{i\delta_{\rm CP}} & 
    c_{12} c_{s3} -s_{12} s_{13} s_{23} e^{i\delta_{\rm CP}}  & 
    c_{13} s_{23} \\
    s_{12}s_{23} - c_{12} s_{13} c_{23} e^{i\delta_{\rm CP}} & 
    -c_{12}s_{23} -s_{12} s_{13} c_{23} e^{i\delta_{\rm CP}}& 
    c_{13} c_{23}
\end{bmatrix}
{\rm U}_{\rm M}
\label{upmnsmatrix}
\end{eqnarray*}

where $s(c)_{12,13,23} \equiv \sin(\cos)\theta_{12,13,23}$ are three mixing angles and {$\delta_{\rm CP}$} is the Dirac CP phase. The matrix 
${\rm U}_{\rm M} \equiv \rm{diag} (e^{i \alpha_1}, e^{i \alpha_2},1)$ is associated with additional CP phases  $\alpha_{1,2}$ in case neutrinos are self-conjugate Majorana particles. However, oscillation experiments are not sensitive to $\alpha_{1,2}$.
 
Recent NuFIT 5.2 with SK atmospheric data (NuFIT)
\cite{Esteban:2020cvm,NuFIT5.2data} 
gives at the 3$\sigma$ level (1dof, $\Delta\chi^2=9$) 
\begin{eqnarray}
    \rm{NO:} &&\theta_{13} \in (8.23^{\circ}, 8.91^{\circ}), \theta_{12} \in (31.31^{\circ}, 35.74^{\circ}), \theta_{23} \in (39.7^{\circ}, 51.0^{\circ}), \nonumber \\
    && \delta_{\rm CP} \in (144^{\circ}, 350^{\circ}), \label{eq:NO}\\
    \rm{IO:} && \theta_{13} \in (8.23^{\circ}, 8.94^{\circ}), \theta_{12} \in (31.31^{\circ}, 35.74^{\circ}), \theta_{23} \in (39.9^{\circ}, 51.5^{\circ}), \nonumber \\
    &&\delta_{\rm CP} \in (194^{\circ}, 344^{\circ}). \label{eq:IO}
\end{eqnarray}
For a recent review on global fits, see \cite{Chauhan:2022gkz}. 
With experimental improvements, a further increase in precision is expected in the coming years,   
see Fig.~1 in \cite{Song:2020nfh}.
  
\begin{figure}[!htb]
    \centering
     \includegraphics[width=0.49\textwidth]{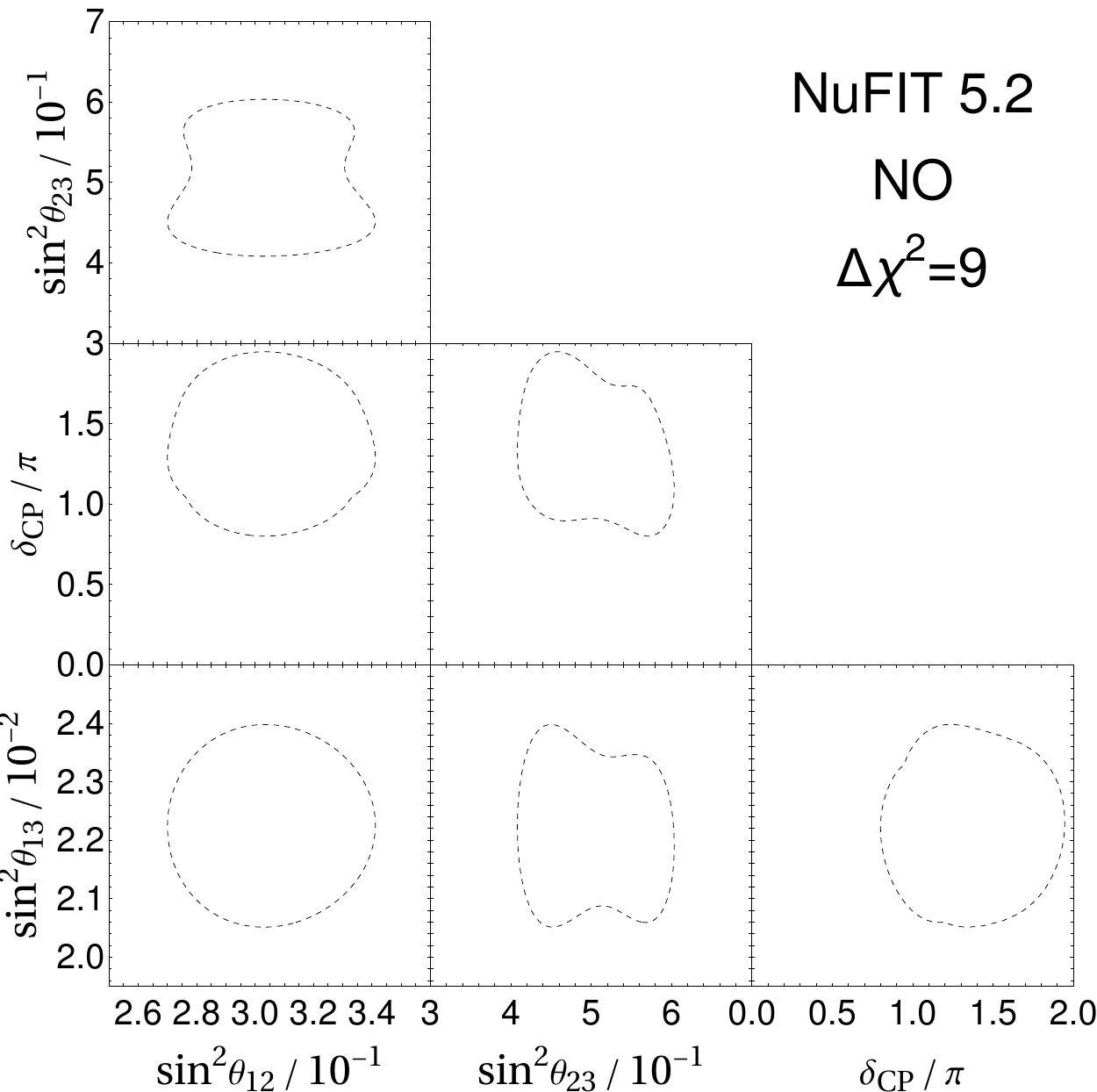}
    \includegraphics[width=0.49\textwidth]{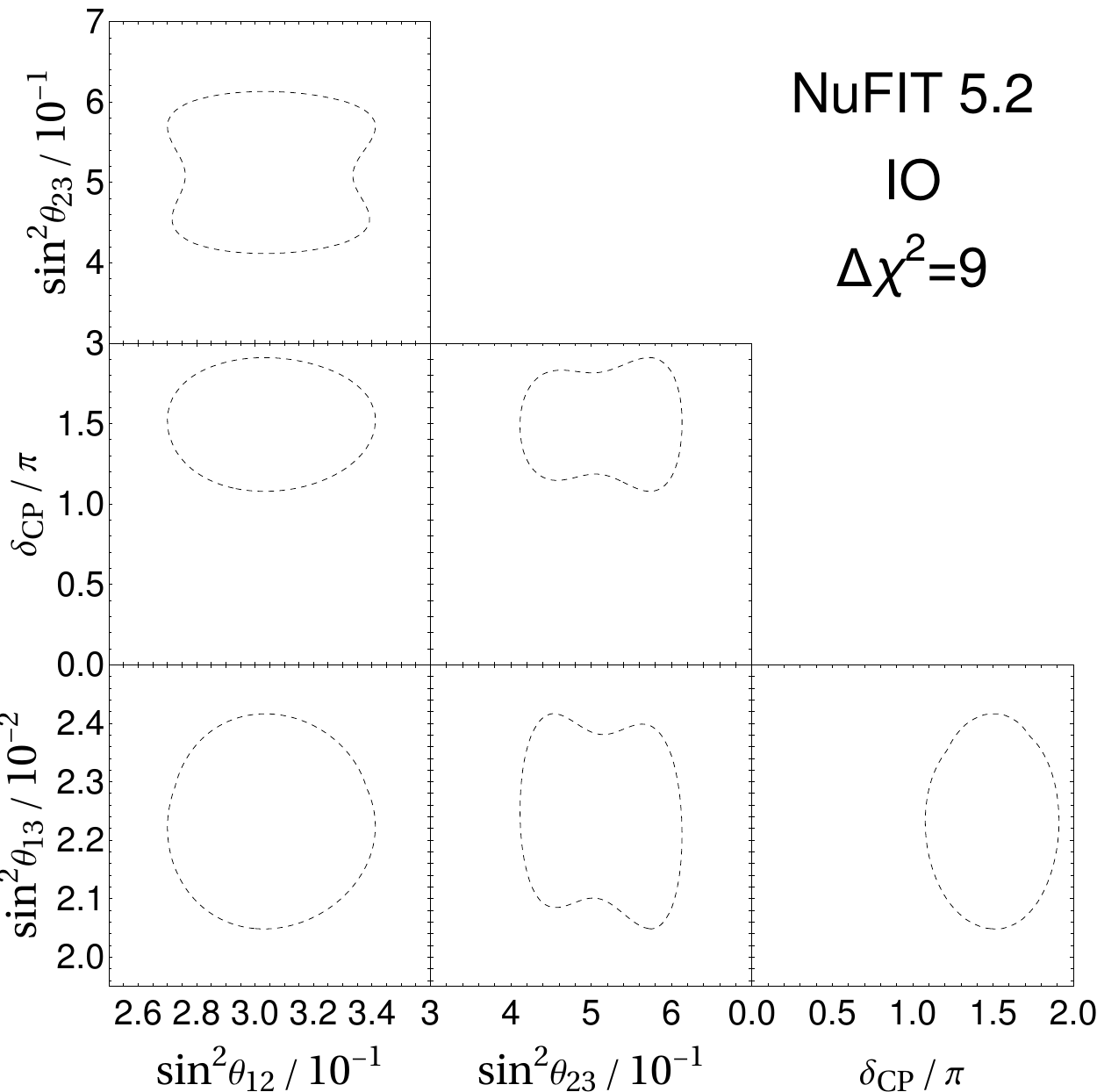}
     \caption{  
     Sample correlations of oscillation parameters at 
     $\Delta\chi^2=9$
     for NO and IO mass ordering. Plots based on NuFIT data~\cite{NuFIT5.2data}. 
     }
    \label{fig:1}
\end{figure}

In this proceedings, we show how correlations among neutrino parameters can be used to constrain realistic mixing schemes. The correlations for specific sets of oscillation parameters are provided by NuFIT 
\cite{Esteban:2020cvm,NuFIT5.2data} or de Salas et al. \cite{deSalas:2020pgw} in the form of $\Delta\chi^2$ tables. Here we use the most recent NuFIT 5.2 data sets (with SK atmospheric data). In Fig.~\ref{fig:1} we show sample correlations among neutrino mixing parameters for both normal (NO) and inverted (IO) mass ordering at $\Delta\chi^2=9$. 

 \FloatBarrier

Origin of the observed pattern of neutrino mixing still is one of the most fundamental challenges in neutrino physics. Theories based on non-Abelian discrete flavor symmetries are among the most elegant frameworks which proposes various mixing schemes. With the accurate measurement of $\theta_{13}$ in 2012 by Daya Bay~\cite{An:2012eh} and Reno~\cite{Ahn:2012nd}, trimaximal (TM$_1$ and TM$_2$) \cite{Albright:2010ap} mixing scheme stands out as a plausible explanation for the lepton mixing matrix. In this work we attempt to further constraint the $\rm{TM}_1$ and $\rm{TM}_2$ mixing predictions, with the correlations obtained from the observed neutrino oscillation data at $\Delta\chi^2{\approx}6.18/9/11.83$ levels for both NO and IO (see Fig.~\ref{fig:1}). 
 
\section{Constraining $\rm{TM}_1$ and $\rm{TM}_2$ predictions}
\label{section_tm}
In the $\rm{TM}_1$ and $\rm{TM}_2$ mixing schemes, trimaximal mixing matrix has the following structure 
\[
\label{u_tm1tm2}
\begin{array}{cc}
|\rm{U_{{TM}_1}}|=	
					\begin{bmatrix}
						\frac{2}{\sqrt{6}} &* &*\\
						\frac{1}{\sqrt{6}} &* &*\\
						\frac{1}{\sqrt{6}} &* &*
					\end{bmatrix},
     &
      |\rm{U_{{ TM}_2}}|=	
					\begin{bmatrix}
						* &\frac{1}{\sqrt{3}} &*\\
						* &\frac{1}{\sqrt{3}} &*\\
						* &\frac{1}{\sqrt{3}} &*
					\end{bmatrix}. 
\end{array}
\]
Comparing the corresponding elements of the first (second) column of $\rm{U_{PMNS}}$ and $\rm{U_{{TM}_1}}$ ($\rm{U_{{TM}_2}}$), relations between oscillation parameters can be derived (see e.g. \cite{deMedeirosVarzielas:2012apl,Chauhan:2023faf}).
The relation between $s_{12}^2$ and $s_{13}^2$ reads
\begin{equation}
     \label{eq:tms12s13}
     \mbox{TM}_1 :  s_{12}^2 = \frac{1-3 s_{13}^2}{3 - 3 s_{13}^2}, 
    ~~~~~~~~~
     \mbox{TM}_2 : s_{12}^2 = \frac{1}{3 - 3 s_{13}^2}.
\end{equation}

\noindent Similarly, the relation between $\delta_{\rm CP}$ and $s_{13}^2$ and $s_{23}^2$ can be written as 
\begin{eqnarray}
     \label{eq:tmdcps23s13}
     \nonumber
     \mbox{TM}_1 :  \cos \delta_{\rm CP} = \frac{
    (1-5s_{13}^2)(2s_{23}^2-1)
    }{
    4 s_{13}s_{23}\sqrt{2(1-3s_{13}^2)(1-s_{23}^2)}
    }, \\
     \mbox{TM}_2 : \cos \delta_{\rm CP} = -\frac{
    (2-4s_{13}^2)(2s_{23}^2-1)
    }{
    4 s_{13}s_{23}\sqrt{(2-3s_{13}^2)(1-s_{23}^2)}
    }.
\end{eqnarray}
\noindent 
Although Eq.~(\ref{eq:tmdcps23s13}) explicitly involves only $\delta_{\rm CP}$, $s_{13}^2$ and $s_{23}^2$, it has a hidden dependence on $s_{12}^2$ via Eq.~(\ref{eq:tms12s13}).
Hence to constrain the TM$_{1,2}$ predictions we use all possible correlations, such as,  $s_{13}^2$ \textit{vs} $s_{12}^2$, 
 $s_{13}^2$ \textit{vs} $s_{23}^2$, 
 $s_{12}^2$ \textit{vs} $s_{23}^2$,
 $s_{13}^2$ \textit{vs} $\delta_{\rm CP}$,
 $s_{12}^2$ \textit{vs} $\delta_{\rm CP}$,
 $s_{23}^2$ \textit{vs} $\delta_{\rm CP}$ obtained from global analysis of neutrino oscillation data (see Fig.~\ref{fig:1}).
\begin{figure}[!htb]
    \centering
    \includegraphics[width=0.47\textwidth]{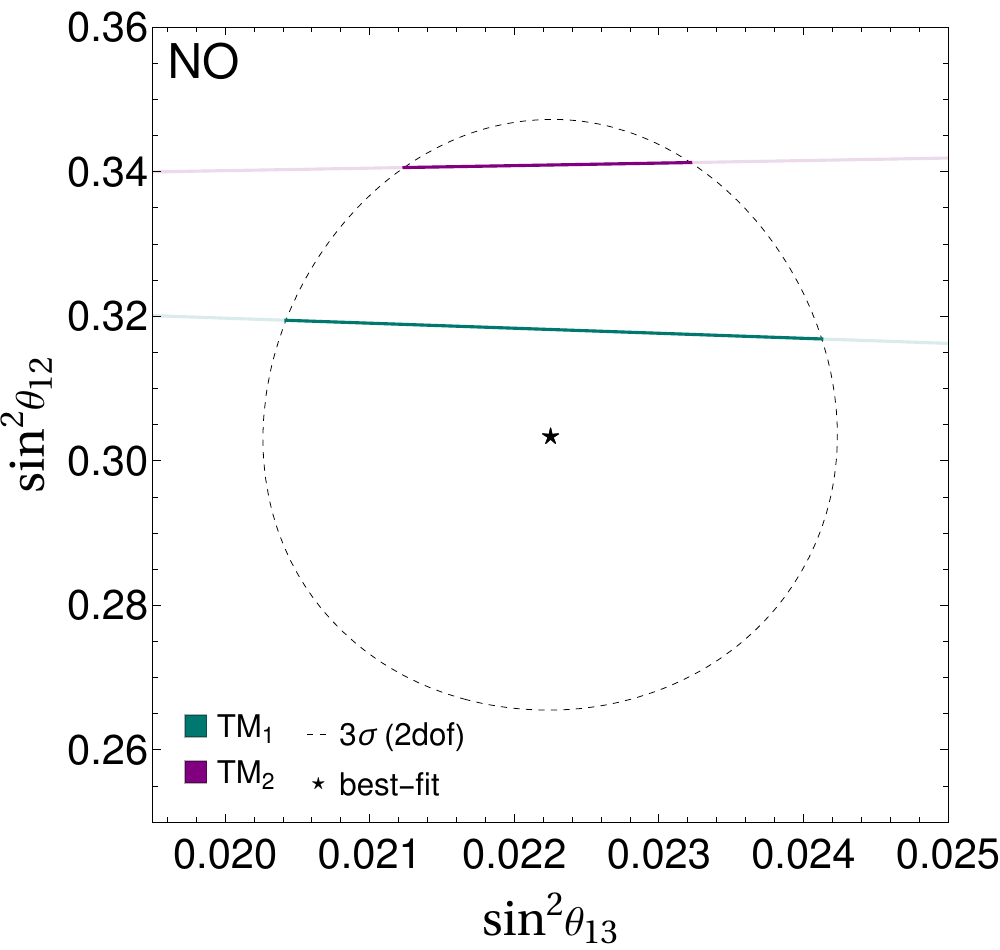}
    \includegraphics[width=0.47\textwidth]{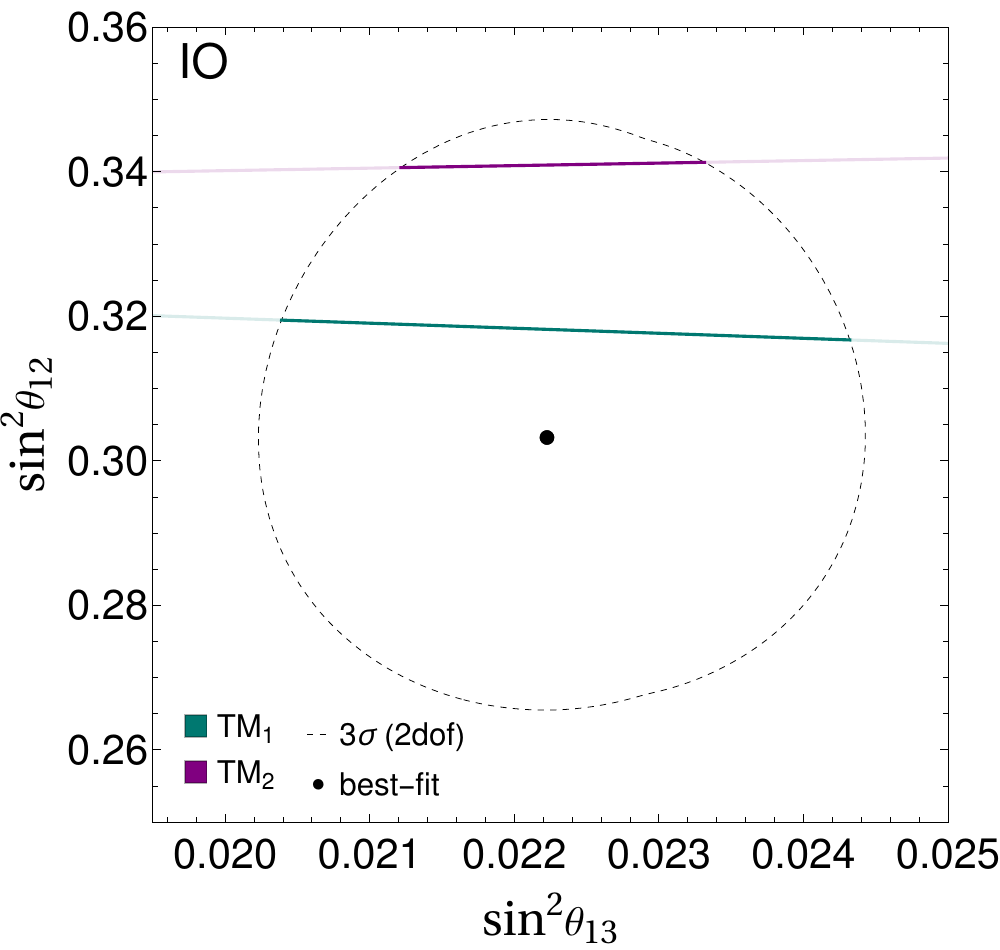}
     \caption{$s_{12}^2$ plotted against $s_{13}^2$ 
      for $\rm{TM}_{1,2}$ mixing. See text for details. 
     }
    \label{fig:tms12s133sigma}
\end{figure}
\begin{figure}[!htb]
    \centering
    \includegraphics[width=0.47\textwidth]{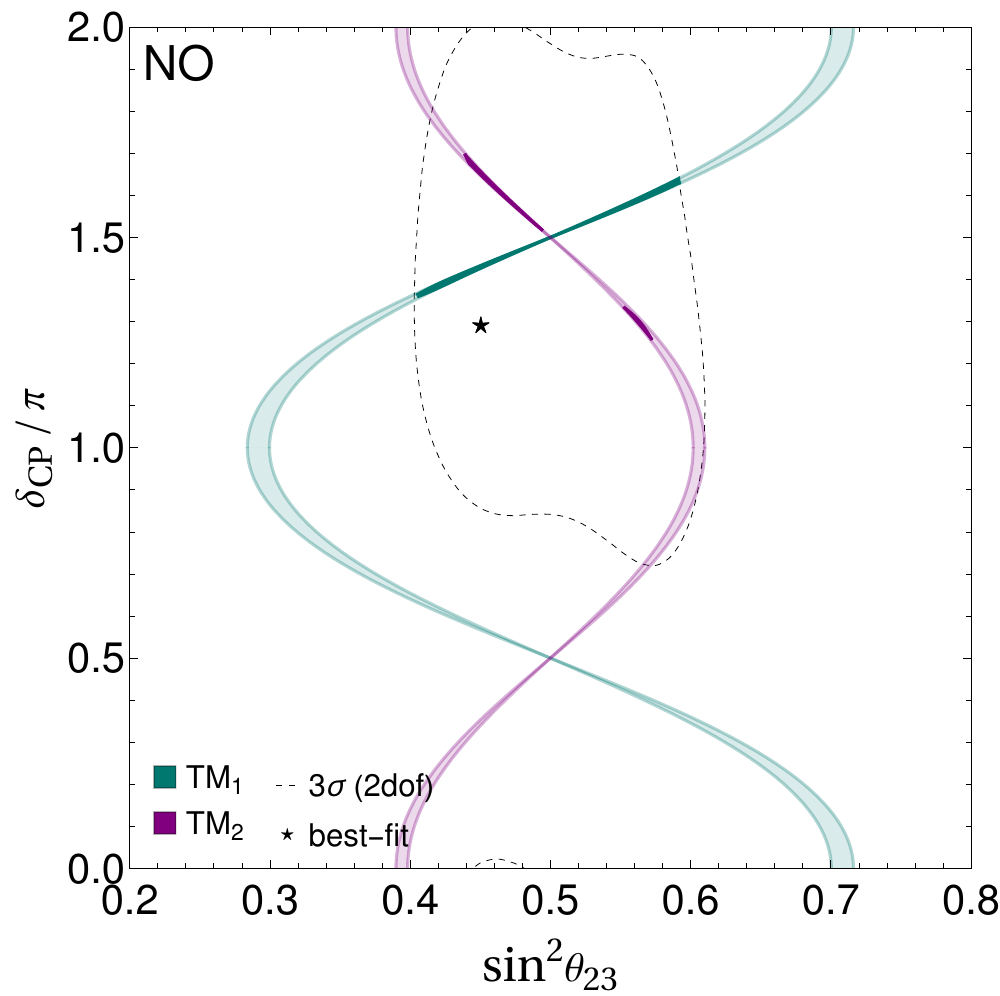}
    \includegraphics[width=0.47\textwidth]{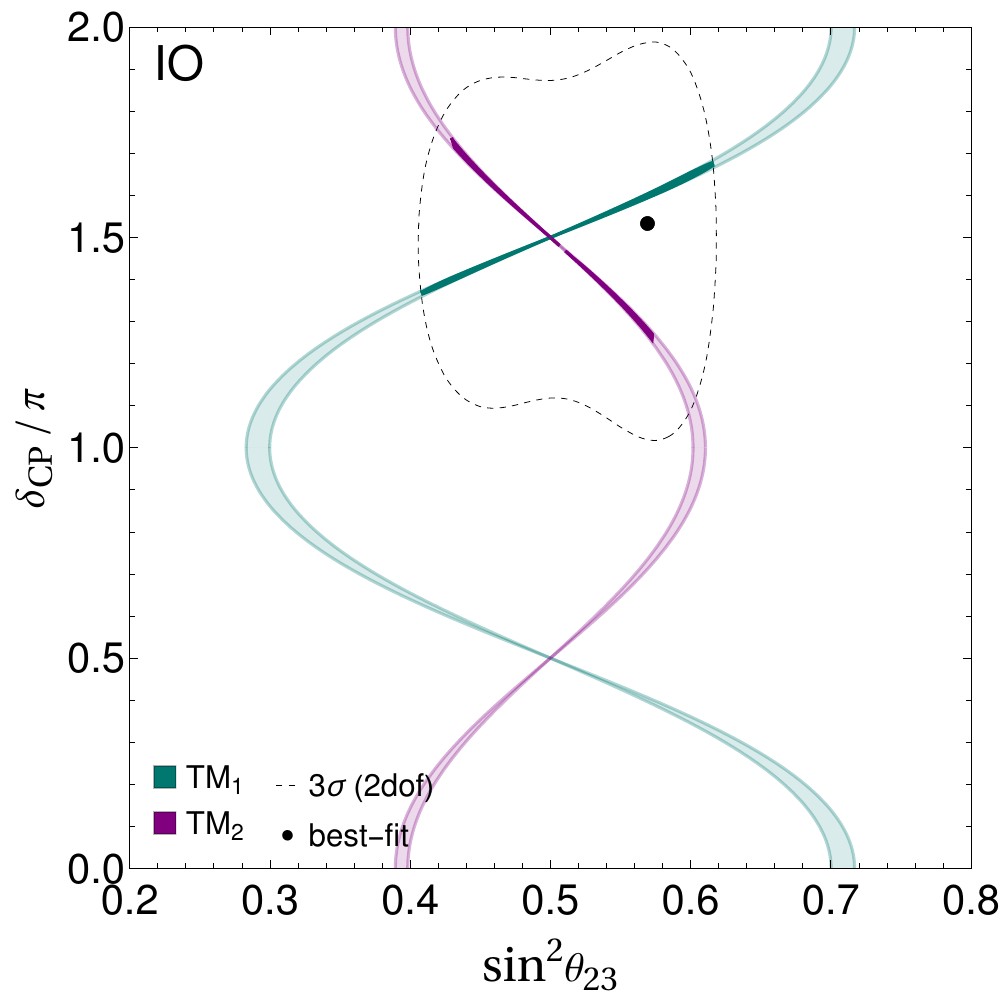}
     \caption{
     $\delta_{\rm CP}$ plotted against $s_{23}^2$ 
     for $\rm{TM}_{1,2}$ mixing. See text for details. 
     }
    \label{fig:tms23dcp3sigma}
\end{figure} 
In Figs.~\ref{fig:tms12s133sigma} and \ref{fig:tms23dcp3sigma} we have plotted dependence of mixing parameters in the $s_{12}^2-s_{13}^2$ and $\delta_{\rm CP} - s_{23}^2$  planes (Eqs.~(\ref{eq:tms12s13}),(\ref{eq:tmdcps23s13})) as theorized by the TM$_{1}$ (light green shaded region) and TM$_{2}$ (light purple shaded region) mixing schemes for both NO (left panel) and IO (right panel). Here the dashed lines imply 3$\sigma$ (2dof) allowed regions by NuFIT~\cite{NuFIT5.2data} and the best-fit values are given by $\star$ (NO), $\bullet$ (IO). The darker shades in Fig.~\ref{fig:tms12s133sigma} (inside the dashed lines) imply the allowed 3$\sigma$ (2dof) region for TM$_{1,2}$ mixings. Using the ranges given in Fig.~\ref{fig:tms12s133sigma} and correlations from NuFIT (see Fig.~\ref{fig:1}), we can further constrain the predictions of TM$_{1,2}$ mixing schemes. For example, the darker-shaded region in Fig.~\ref{fig:tms23dcp3sigma} represents the allowed regions in the $\delta_{\rm CP} - s_{23}^2$ plane significantly constraining the  theoretical prediction for the TM$_2$ mixing. Similar constraints can also be imposed on  $\delta_{\rm CP}-s_{13}^2$ and $s_{23}^2-s_{13}^2$ planes.  
In Fig.~\ref{fig:tms12s133sigma} we can find that $\theta_{12}$ is almost horizontal, and it constraints $\theta_{13}$ strongly for TM$_2$ compared to TM$_1$. These stringent constraints affect the allowed ranges of $\delta_{\rm CP}$ and $\theta_{23}$ in Fig.~\ref{fig:tms23dcp3sigma}. Therefore, finally, complying  with allowed regions of neutrino oscillation parameters (Fig.~\ref{fig:1}) and their one-dimensional projections in $s^2_{13}-s^2_{12}$ (Fig.~\ref{fig:tms12s133sigma}),  $\delta_{\rm CP} - s_{23}^2$ (Fig.~\ref{fig:tms23dcp3sigma}), $\delta_{\rm CP}-s_{13}^2$ and $s_{23}^2-s_{13}^2$ planes, we have summarized the final prediction (on $\theta_{12,23,13}$ and $\delta_{\rm CP}$) for TM$_{1,2}$ mixing schemes in Tab.~\ref{tab:constr_data_tm1and2} for both mass orderings. The results in this table should be compared with `model independent' results of a global fit in Eqs.~(\ref{eq:NO}) and (\ref{eq:IO}).

\begin{table}[h!]
 \centering
    \scriptsize
    \begin{tabular}{c c c c c}
    \hline \hline
         \multicolumn{5}{c}{$\rm{TM}_1$ mixing}
            \\
            \hline
             Parameter &  Ordering &
        $\Delta\chi^2\approx6.18$
        &  
        $\Delta\chi^2=9$
        &  
        $\Delta\chi^2\approx11.83$
        \\
         \hline 
          $\theta_{13}/^{\circ}$
        & NO
        &$8.33 - 8.83$
        &$8.27 - 8.89$
        &$8.21 - 8.94$
        \\
        & IO
        &$8.33 - 8.85$
        &$8.26 - 8.92$
        &$8.21 - 8.98$
        \\
        \hline
        $\theta_{12}/^{\circ}$
        & NO
        &$34.28 - 34.39$
        &$34.26 - 34.41$
        &$34.25 - 34.42$
        \\
        & IO
        &$34.27 - 34.39$
        &$34.26 - 34.41$
        &$34.24 - 34.42$
        \\
        \hline
         $\theta_{23}/^{\circ}$
        & NO
        &$40.3 - 45.6$
        &$39.8 - 49.5$
        &$39.5 - 50.3$
        \\
        & IO
        &$40.5 - 51.0$
        &$40.1 - 51.4$
        &$39.7 - 51.7$
        \\
        \hline
         $\delta_{\rm CP}/^{\circ}$
        & NO
        &$248.6 - 272.4$
        &$246.4 - 290.8$
        &$244.7 - 295.2$
        \\
        & IO
        &$250.0 - 297.7$
        &$247.7 - 300.1$
        &$245.8 - 302.0$
        \\
    \hline \hline
         \multicolumn{5}{c}{$\rm{TM}_2$ mixing}
            \\
            \hline
             Parameter &  Ordering &
        $\Delta\chi^2\approx6.18$
        &  
        $\Delta\chi^2=9$
        &  
        $\Delta\chi^2\approx11.83$
        \\
         \hline 
          $\theta_{13}/^{\circ}$
         & NO
        & $\times$ 
        &$8.53 - 8.62$
        &$8.38 - 8.77$
        \\
        & IO
        & $\times$
        &$8.53 - 8.61$
        &$8.37 - 8.79$
        \\
        \hline
        $\theta_{12}/^{\circ}$
        & NO
        & $\times$ 
        &$35.72 - 35.73$
        &$35.70 - 35.75$
        \\
        & IO
        & $\times$ 
        &$35.72 - 35.73$
        &$35.70 - 35.75$
        \\
        \hline
         $\theta_{23}/^{\circ}$
        & NO
        & $\times$ 
        & $\times$ 
        &$41.5 - 44.7  ~\&~ 48.0 - 49.2$
        \\
        & IO
        & $\times$ 
        & $\times$ 
        &$40.9 - 45.4  ~\&~ 45.6 - 49.2$
        \\
        \hline
         $\delta_{\rm CP}/^{\circ}$
        & NO
        & $\times$ 
        & $\times$ 
        &$226.6 - 239.9  ~\&~ 273.3 - 305.3$
        \\
        & IO
        & $\times$ 
        & $\times$ 
        &$226.2 - 263.9 ~\&~ 266.9 - 312.0$
        \\
        \hline \hline
    \end{tabular}
    \caption{Constraints on $\rm{TM}_1$ and TM$_2$ mixing scenarios obtained using correlations among neutrino oscillation parameters inferred from experimental data.  $\Delta\chi^2{\approx}6.18/9/11.83$ corresponds to $2\sigma$/$2.54\sigma$/$3\sigma$ (2dof). 
    The symbol "$\times$" indicates that there are no allowed solutions for TM$_2$ mixing.
    }
\label{tab:constr_data_tm1and2}
\end{table}
 
\section{Conclusions}
We show how correlations obtained from neutrino oscillation data affect constraints on the $\rm{TM}_1$ and $\rm{TM}_2$ mixings.   
Within  allowed regions, the TM$_2$ mixing scheme is most constrained.
The outlined here procedure can be applied to analysis of other neutrino mixing models which predict analytic relations among oscillation parameters. 

\section*{Acknowledgements}
We would like to thank Janusz Gluza for discussions and remarks. The research has been supported by the Polish National Science Center (NCN) under grant 2020/37/B/ST2/02371.

\bibliography{bibliography}

\end{document}